\documentstyle[twocolumn,prl,aps,epsfig,subeqn]{revtex}
\def\pr#1#2#3{Phys.~Rev~{\bf #1},\ #2\ (#3)}
\def\pra#1#2#3{Phys.~Rev.~A~{\bf #1},\ #2\ (#3)}
\def\prl#1#2#3{Phys.~Rev.~Lett.~{\bf #1},\ #2\ (#3)}
\def\sci#1#2#3{Science~{\bf #1},\ #2\ (#3)}
\def\nat#1#2#3{Nature~{\bf #1},\ #2\ (#3)}
\def\hphi{\hat{\Phi}}
\def\psiab{\hat{\Psi}_{AB}}
\def\psic{\hat{\Psi}_C}
\def\psia{\hat{\Psi}_A}
\def\psibc{\hat{\Psi}_{BC}}
\def\h{\hat{H}}
\def\oo{\Omega_1}
\def\ot{\Omega_2}
\def\bfr{{\bf r}}
\def\ddt{\frac{d}{dt}}
\def\adamo#1#2#3{Adv.~At.~Mol.~Opt.~Phys. {\bf #1},\ #2\ (#3)}
\newcommand{\beq}{\begin{equation}}
\newcommand{\eeq}{\end{equation}}

\begin{document}
\flushbottom \draft
\title{Bose-enhanced chemistry: Amplification of selectivity in the
dissociation of molecular Bose-Einstein condensates}
\author{M. G. Moore and A. Vardi}
\address{ITAMP, Harvard-Smithsonian Center for Astrophysics, Cambridge,
MA 02138 \\ (\today)
\\ \medskip}\author{\small\parbox{14.2cm}{\small\hspace*{3mm}
We study the photodissociation chemistry of a quantum degenerate gas of bosonic triatomic
$ABC$ molecules, assuming two open rearrangement channels ($AB+C$ or $A+BC$). The equations
of motion are equivalent to those of a parametric multimode laser, resulting in an
exponential buildup of macroscopic mode populations. By exponentially amplifying a small
differential in the single-particle rate-coefficients, Bose stimulation leads to a nearly
complete selectivity of the collective $N$-body process, indicating a novel type of
ultra-selective quantum degenerate chemistry.\\
\\[3pt]PACS numbers: 03.75.Fi, 03.75.Be, 03.75.-b }}
\maketitle
\maketitle \narrowtext

A significant effort is currently underway with the goal of producing a stable
Bose-Einstein condensate in a vapor of molecules. Several groups are pursuing schemes in
which atomic BECs are converted to molecular BECs via photoassociation \cite{wynar}
and/or Feshbach resonance \cite{inouyeo,stenger,cornish}. Progress is also being made
towards forming a molecular BEC by direct evaporation of molecules in a magnetic trap. In
this case sympathetic buffer-gas cooling is used to load the trap \cite{weinstein}, as
opposed to the optical techniques used in the alkali-metal atomic vapors.

Once the technical problems are solved, the realization of molecular condensates will 
open a new field of quantum-degenerate coherent chemistry.
The most striking new feature of the resulting molecular dynamics would be the
unprecedented role of nonlinear collective effects, i.e. Bose enhancement and Pauli
blocking, induced by the $N$-body quantum statistics. The probability of any single boson
to undergo a transition is enhanced by the number of identical particles occupying the
final state, whereas fermions would be subject to Pauli blocking of the output channels.
Thus any process which retains a critical phase-space density would greatly differ from
that of the isolated single-particle process which dominates 'thermal' gas-phase chemistry
\cite{collective}.

The role of Bose stimulation in the association and dissociation of a molecular BEC has
already been studied theoretically in the regime where a single atomic mode is coupled to a
single molecular mode \cite{timmermans,vanabeelen,javanainen,heinzen}. In this situation
the entire condensate is theoretically predicted to undergo large amplitude coherent
oscillations between atoms and molecules . The Bose-enhanced oscillation frequency is
$\sqrt{N}\Omega$, where $\Omega$ is the single-particle atom-molecule coupling frequency
and $N$ is the total number of condensate particles. While this prediction is significantly
modified by the inclusion of quantum-field effects \cite{holland,goral,hope,vardi}, the
collective evolution remains strikingly different from the single-pair (non-degenerate)
dynamics.

The purpose of this Letter is to extend these ideas to the case where more complicated
molecules are dissociated into a multiplicity of output channels. In such multi-channel 
quantum-degenerate processes there are two important competing effects: Bose stimulation,
which leads to exponential growth of the output channel populations, and condensate
depletion, in which the finite number of initial molecules leads to competition between
output channels. If certain channels have a slight advantage in coupling strength, 
they will deplete the condensate before other channels attain a macroscopic 
population. Related effects have been experimentally observed in the phenomenon of 
superradiant Rayleigh scattering of laser light from an elongated 
BEC \cite{inouyet,moore}, indicating that there are no overwhelming obstacles to 
obtaining winner-takes-all scenarios in BEC systems.
\begin{figure}
\begin{center}
\epsfig{file=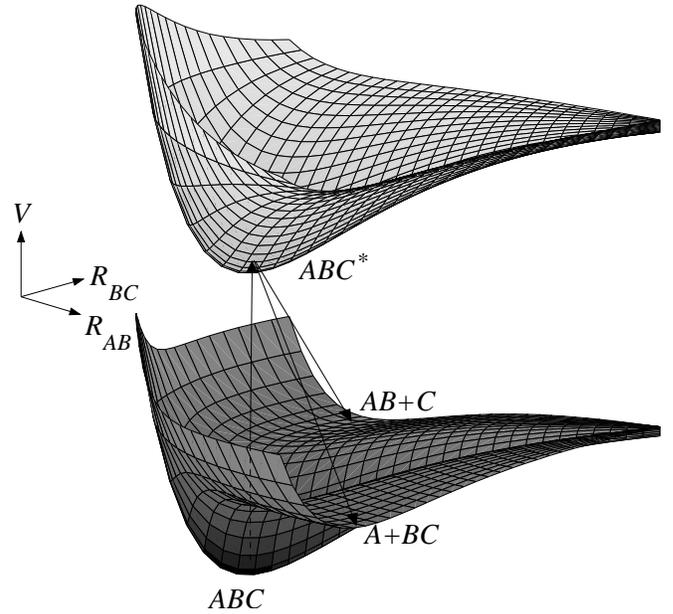,width=\columnwidth}
\end{center}
\caption{Stimulated Raman Photodissociation of $ABC$ molecules} \label{f1}
\end{figure}

Studying the simplest example of the dissociation of a triatomic ABC molecule 
into AB+C or A+BC pairs, we show that the combined effects of Bose stimulation and 
competition between output channels for a finite initial number of molecules leads 
to an {\it amplification of selectivity} which can dramatically enhance any 
non-degenerate control scheme. Relatively small controllability of the single-particle 
process translates to near complete controllability of the Bose-stimulated $N$-body 
process. Thus, collective amplification and mode competition effects can be employed 
to achieve the central goal of the fields of optimal- and 
coherent-control \cite{shapiro,rice,rabitz}.

We consider a BEC of triatomic $ABC$ molecules. The ground-state molecules are resonantly
coupled by a stimulated Raman transition to two open dissociation channels: $AB+C$ ($AB$
molecules and $C$ atoms) and $A+BC$ ($A$ atoms and $BC$ molecules), as depicted in Fig. 1.
After adiabatic elimination of the intermediate $ABC^*$ state the $N$-body second-quantized
Hamiltonian in the interaction representation and the rotating wave approximation, reads
\beq
    \h=\frac{\hbar}{2}\hphi(\bfr)\left[\oo\psiab^\dagger(\bfr)\psic^\dagger(\bfr)
    +\ot\psia^\dagger(\bfr)\psibc^\dagger(\bfr)\right]+h.c.
\label{ham}
\eeq
where $\oo,\ot$ are the coupling frequencies into the $AB+C$ and $A+BC$
dissociation channels, respectively and $\hphi$, $\psiab$, $\psic$,
$\psia$, $\psibc$ are the bosonic field operators for $ABC$, $AB$, $C$,
$A$, $BC$, respectively.

Using the Hamiltonian (\ref{ham}), the Heisenberg equations of motion for these operators
are
\noindent
\begin{subequations}
\begin{eqnarray}
    i\ddt\hphi&=&\oo\psiab\psic+\ot\psia\psibc
\label{hpieom}\\
    i\ddt\psiab&=&\oo\hphi\psic^\dagger
\label{psiabeom}\\
    i\ddt\psic^\dagger&=&\oo^*\hphi^\dagger\psiab
\label{psiceom}\\
    i\ddt\psia&=&\ot\hphi\psibc^\dagger
\label{psiaeom}\\
    i\ddt\psibc^\dagger&=&\ot^*\hphi^\dagger\psia
\label{psibceom}
\end{eqnarray}
\label{eom}
\end{subequations}

Equations (\ref{eom}) are closely related to the quantum equations of motion of a multimode
laser \cite{mml}, in that the $AB+C$ dissociation channel (Eqs.
(\ref{psiabeom},\ref{psiceom})) is only coupled to the $A+BC$ channel (Eqs.
(\ref{psiaeom},\ref{psibceom})) via the molecular reservoir $\hphi$. In order to
demonstrate the collective amplification of selectivity, we use a five-mode model,
neglecting the spatial variation of the various fields. This approximation, which assumes
there is only one available state per specie, is valid for a zero temperature $ABC$
condensate at exact resonance with the dissociation threshold of the $AB+C$ and $A+BC$
channels. The field operators of Eqs. (\ref{eom}) are thus taken to represent the particle
creation and annihilation operators of the respective modes. In the undepleted pump
approximation \cite{goral,vardi}, the intense $\hphi$ field is replaced by a constant
c-number $\Phi\sim\sqrt{N}$. Assuming without loss of generality that $\oo=\oo^*$ 
and $\ot=\ot^*$, the initial time evolution of the operators $\psiab,\psic,\psia$, 
and $\psibc$ is thus given as
\begin{subequations}
\begin{eqnarray}
    \psiab(t)&=&\psiab(0)\cosh \omega_1 t-i\psic^\dagger(0)\sinh \omega_1 t~,\\
    \psic^\dagger(t)&=&\psic^\dagger(0)\cosh\omega_1 t +i\psiab(0)
    \sinh \omega_1 t~,\\
    \psia(t)&=&\psia(0)\cosh \omega_2 t-i\psibc^\dagger(0)\sinh \omega_2 t~,\\
    \psibc^\dagger(t)&=&\psibc^\dagger(0)\cosh\omega_2 t +i\psia(0)
    \sinh \omega_2 t~,
\end{eqnarray}
\label{psisol}
\end{subequations}
where $\omega_1=\sqrt{N}\oo$ and $\omega_2=\sqrt{N}\ot$. Assuming $\langle\psiab(0)\rangle
= \langle\psic(0)\rangle = \langle\psia(0)\rangle = \langle\psibc(0)\rangle = 0$ (no
initial product coherence), the average numbers of particles $\langle n_j\rangle =
\langle\Psi_j^\dagger\Psi_j\rangle$ in the modes $j=AB,C,A,BC$, are given, respectively, as
\begin{subequations}
\begin{eqnarray}
    \langle n_{AB}(t)\rangle&=&
    \langle n_{AB}(0)\rangle\cosh^2\omega_1 t\nonumber\\
    ~&~&+(1+\langle n_{C}(0)\rangle)\sinh^2\omega_1 t~,\\
    \langle n_{C}(t)\rangle&=&
    \langle n_{C}(0)\rangle\cosh^2\omega_1 t\nonumber\\
    ~&~&+(1+\langle n_{AB}(0)\rangle)\sinh^2\omega_1 t~,\\
    \langle n_{A}(t)\rangle&=&
    \langle n_{A}(0)\rangle\cosh^2\omega_2 t\nonumber\\
    ~&~&+(1+\langle n_{BC}(0)\rangle)\sinh^2\omega_2 t~,\\
    \langle n_{BC}(t)\rangle&=&
    \langle n_{BC}(0)\rangle\cosh^2\omega_2 t\nonumber\\
    ~&~&+(1+\langle n_{A}(0)\rangle)\sinh^2\omega_2 t~
\end{eqnarray}
\end{subequations}
Starting with a pure ABC condensate
$\langle n_{AB}(0)\rangle=\langle n_{C}(0)\rangle=
\langle n_{A}(0)\rangle=\langle n_{BC}(0)\rangle=0$,
the average product populations grow as
\begin{subequations}
\begin{eqnarray}
    \langle n_{AB}(t)\rangle=\langle n_{C}(t)\rangle=
    \sinh^2\sqrt{N}\oo t~,\\
    \langle n_{A}(t)\rangle=\langle n_{BC}(t)\rangle=
    \sinh^2\sqrt{N}\ot t~.
\end{eqnarray}
\label{ngrowth}
\end{subequations}
The initial time evolution of Eqs. (\ref{ngrowth}) corresponds to an initial nonexponential
(quadratic rather than linear in time) spontaneous dissociation process, followed by an
exponential stimulated process \cite{vardi}. We note that exactly the same
equations have been routinely
used for over three decades, to describe the phenomenon of parametric superfluorescence
\cite{louisell,harris}, i.e. the parametric amplification of noise photons, in quantum
optics. The feature which produces the striking selectivity of Bose-enhanced chemistry is
that any small difference between the two rate coefficients, $\ot\neq\oo$, is exponentially
amplified by the stimulated dissociation, so that
\beq
N_2(t)/N_1(t)\approx
e^{2\sqrt{N}(\ot-\oo)t}.
\label{cpr}
\eeq

The linearized model of Eqs. (\ref{psisol}) is only valid as long as the population of the
triatomic mode $\langle\hphi^\dagger\hphi\rangle$ is large and the effect of its depletion
on the products population growth is negligible. In order to go beyond this approximation
we solve the full $N$-body problem numerically. Starting with no products and fixing the
total number of particles $N$, the accessible Hilbert space is restricted to Fock states of
the type $|n_{ABC},n_{AB},n_C,n_A,n_{BC}\rangle$ with $n_{AB}=n_C$, $n_A=n_{BC}$ and
$n_{ABC}=N-(n_A+n_C)$. Using these states to represent the Hamiltonian (\ref{ham}) we solve
numerically the $N$-body Schr\"odinger equation,
\begin{eqnarray}
    i\ddt c_{n_1,n_2}=&\oo\sqrt{N-(n_1+n_2)}(n_1+1)c_{n_1+1,n_2}\nonumber\\
    ~&+\oo^*\sqrt{N-(n_1+n_2)+1}n_1 c^*_{n_1-1,n_2}\nonumber\\
    ~&+\ot\sqrt{N-(n_1+n_2)}(n_2+1)c_{n_1,n_2+1}\nonumber\\
    ~&+\ot^*\sqrt{N-(n_1+n_2)+1}n_1 c^*_{n_1,n_2-1}~,
\end{eqnarray}
for the dynamics of the probability amplitudes
\beq
    c_{n_1,n_2}(t)=\langle N-(n_1+n_2),n_1,n_1,n_2,n_2|\Psi\rangle~.
\eeq
\begin{figure}
\begin{center}
\epsfig{file=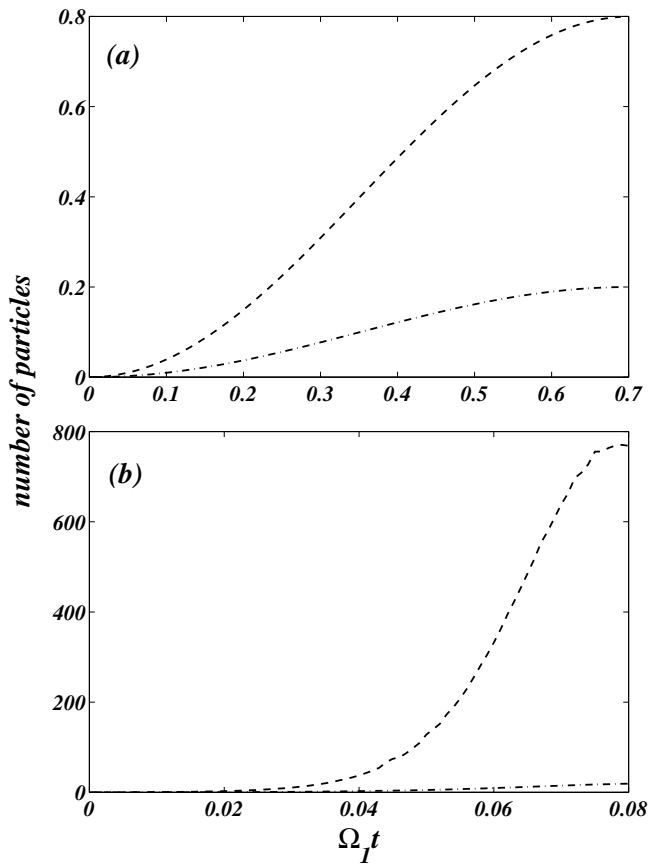,width=\columnwidth}
\end{center}
\caption{Expectation values of the $AB+C$ ($-\cdot-$) and $A+BC$ ($---$) channel
populations as a function of time for (a) $N=1$ and (b) $N=1000$. The coupling rate ratio
is $\ot=2\oo$. } \label{f2}
\end{figure}

In Fig. 2, we plot the expectation values $\langle n_C \rangle$,
$\langle n_A \rangle$, corresponding to the populations in the
$AB+C$ channel and the $A+BC$ channel, respectively, as a function
of time, for $\Omega_2=2\Omega_1$. When $N=1$ (Fig. 2(a)) the ratio between
the two channel populations is, as expected for the single-particle
process $\langle n_A\rangle /\langle n_C\rangle=|\ot/\oo|^2=4$. However,
for $N=1000$ (Fig. 2(b)), the Bose stimulated population ratio is
$\langle n_A\rangle /\langle n_C\rangle>40$. Thus, the reaction outcome
is highly dependent on the total number of particles, becoming more
selective as $N$ increases.
\begin{figure}
\begin{center}
\epsfig{file=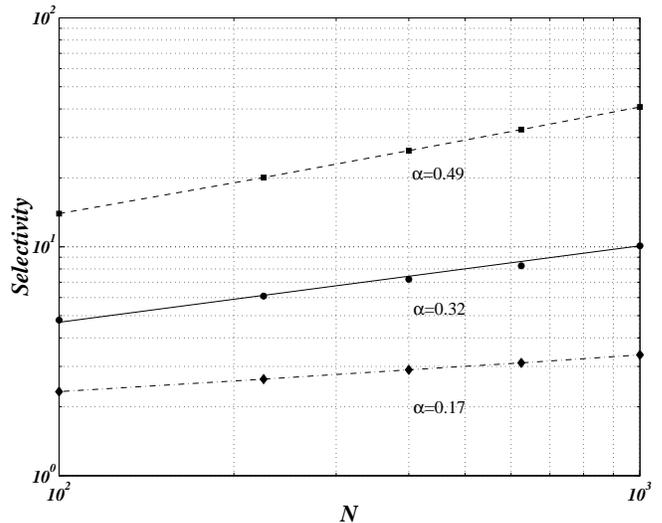,angle=-90,width=\columnwidth}
\end{center}
\caption{Channel population ratio $\langle n_A \rangle / \langle n_C \rangle$ when
${\langle n_A \rangle}={\langle n_A \rangle}_{max}$ as a function of $N$ for $\ot=1.25\oo$
(dash-dot), $\ot=1.5\oo$ (solid), and $\ot=2\oo$ (dashes)} \label{f3}
\end{figure}

Our numerical calculations are limited by available computation power to $N\le 1000$.
However, the number of particles in current BEC experiments in trapped dilute alkali gases
is sometimes as high as $10^7$, suggesting a far greater selectivity enhancement. In 
Fig. 3 we plot the channel population ratio when the leading channel population 
reaches a maximum, as a function of $N$ for various values of $\ot/\oo$. The curves
are linear fits to the numerical data, demonstrating a power-law dependence of the 
selectivity as $\sim N^\alpha$. The power $\alpha$ increases with the rate 
coefficient ratio $\ot/\oo$, as shown in Fig. 4.  

In order to analytically estimate the dependence of the selectivity on the total 
particle number, we take a perturbative approach where we first obtain the time 
$t_{d}$ at which the leading channel depletes the condensate, and then evaluate 
the channel population ratio at $t_{d}$. Assuming a purely
exponential growth, the depletion time is $t_{d}=\log N/2\sqrt{N}\ot$. Substituting into
Eq. (\ref{cpr}), we find that the selectivity is enhanced as $N^\alpha$, where \beq
\alpha={1-\oo/\ot}~, \label{alpha} \eeq so that $\log[N_2(t_d)/N_1(t_d)]\approx \alpha\log
N$. This result is confirmed by the numerical calculations in Fig. 3, giving $\alpha$
values of 0.17, 0.32, and 0.49 for $\ot/\oo=1.25$, $\ot/\oo=1.5$, and $\ot/\oo=2$, in good
agreement with the predicted values of 0.2, 0.33, and 0.5, respectively. Extrapolating to
$N=10^6$, we find population ratios that are greater than $13$, $100$, and $1200$ for
$\ot/\oo=1.25$, $\ot/\oo=1.5$, and $\ot/\oo=2$, respectively (as compared to single
particle ratios of 1.56, 2.25, and 4, respectively).

The dependence of the selectivity amplification constant $\alpha$ on the rate coefficient
ratio $\ot/\oo$ is plotted in Fig. 4. The solid line depicts the estimate of Eq.
(\ref{alpha}). At the limit of a high rate coefficient ratio the selectivity is enhanced as
$N$, implying that all the particles would dissociate into the preferred channel. We note
that since the linearized model underestimates the condensate depletion time, the numerical
values of $\alpha$ are consistently lower than the theoretical estimate of Eq. (\ref{alpha}).
\begin{figure}
\begin{center}
\epsfig{file=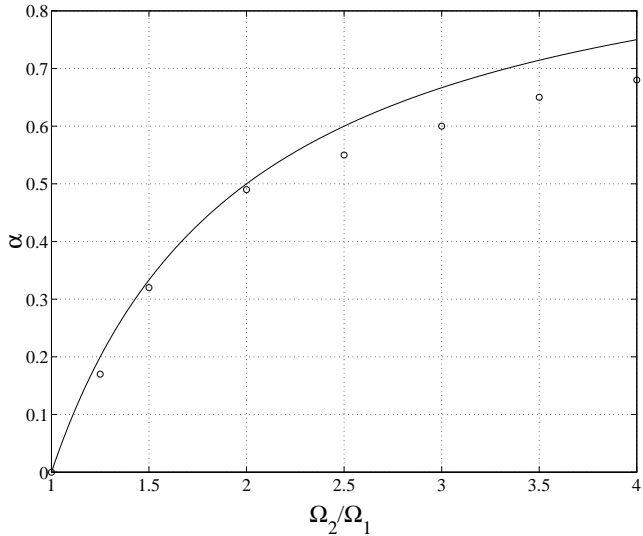,angle=-90,width=\columnwidth}
\end{center}
\caption{The variation of the selectivity amplification constant $\alpha$ with the rate
coefficient ratio $\ot/\oo$. The solid line corresponds to the estimate of Eq.
(\ref{alpha}), whereas the circles are obtained from the numerical quantum solutions}
\label{f4}
\end{figure}

While the five-mode model is sufficient to produce Bose-enhanced gain, it neglects various
spatial effects. In  particular, channel losses due to the relative translational
motion of the products are not accounted for under the assumption that energy
conservation forces the reaction products to be at rest. In fact, by imparting additional
momentum via the laser fields, such losses could be used to increase selectivity by setting
a threshold for amplification. The population in each rearrangement channel would only be
stimulated if the respective pump rate $\sqrt{N}\Omega_j$ is faster than the channel
depletion. Thus, if the motion in the $AB+C$ channel is faster than in the $A+BC$ channel
it will be possible to tune the laser intensity so that only one pathway is Bose
stimulated. In particular, if $A$ and $B$ are atoms of similar mass $m_A\sim m_B$, and $C$
is a much lighter atom $m_C\ll m_A,m_B$, the $A+BC$ channel will be enhanced at much lower
pump intensities than the $AB+C$ channel. Consequently, selectivity could be attained even
for $\oo=\ot$, an issue currently under active investigation.

By drawing an analogy from the phenomena of BEC superradiance \cite{inouyet,moore} to
multi-channel reactive scattering in molecular condensates, we have shown that the
selectivity of reactive processes carried out in a quantum-degenerate gas, can be greatly
amplified by collective enhancement effects. This selectivity is the result of interplay
between bose enhancement, which leads to exponential growth, and competition between modes
for a finite number of initial molecules. The effect was demonstrated using the
simple model system of two-channel photodissociation, where due to a small difference
between the two channel rate coefficients, population was shown to be directed by
the $N$-body stimulated process into the favored channel, approaching a nearly complete
selectivity in the limit of large $N$. In the quantum degenerate regime, it is therefore
only necessary to exert slight control over the reaction rates in order to achieve
near-total control over the reaction output. Future work would address the question 
of complementary Pauli-blocking effects when a molecular BEC is dissociated into 
fermionic constituents. 

We are grateful to James R. Anglin for valuable discussions. This
work was supported by the National Science Foundation through a
grant from the Institute for Theoretical Atomic and Molecular
Physics at Harvard University and the Smithsonian Astrophysical
Observatory.


\begin{thebibliography}{99}
\bibitem{wynar}
R. Wynar {\it et al.}, \sci{287}{1016}{2000}.

\bibitem{inouyeo}
S. Inouye {\it et al.}, \nat{392}{151}{1998}.

\bibitem{stenger}
J. Stenger {\it et al.},\prl{82}{4569}{1999}.

\bibitem{cornish}
S. L. Cornish {\it et al.}, \prl{82}{1795}{2000}.

\bibitem{weinstein}
J. Weinstein {\it et al.}, \nat{395}{148}{1998}.

\bibitem{collective}
It should be noted that collective enhancement and suppression can still occur in the
absence of quantum degeneracy. In general, however, collective states are difficult to
prepare and/or decohere rapidly.

\bibitem{timmermans}
E. Timmermans {\it et al.}, \prl{83}{2691}{1999}.

\bibitem{vanabeelen}
F. A. van Abeelen and B. J. Verhaar, \prl{83}{1550}{1999}.

\bibitem{javanainen}
J. Javanainen and M. Mackie, \pra{59}{R3186}{1999}.

\bibitem{heinzen}
D. J. Heinzen {\it et al.}, \prl{84}{5029}{2000}.

\bibitem{holland}
M. Holland {\it et al.}, \prl{86}{1915}{2001}.

\bibitem{goral}
K. G\'oral {\it et al.}, \prl{86}{1397}{2001}.

\bibitem{hope}
J. J. Hope and M. K. Olsen, \prl{86}{3220}{2001}.

\bibitem{vardi}
A. Vardi {\it et al.}, \pra{64}{063611}{2001}.


\bibitem{inouyet}
S. Inouye {\it et al.}, \sci{285}{571}{1999}.

\bibitem{moore}
M. G. Moore and P. Meystre, \prl{83}{5202}{1999}.



\bibitem{shapiro}
M. Shapiro and P. Brummer, \adamo{42}{287}{2000}.

\bibitem{rice}
S. A. Rice and M. Zhao, {\it Optical Control of Molecular Dynamics}
(Wiely, New York, 2000).

\bibitem{rabitz}
H. Rabitz {\it et al.}, \sci{288}{824}{2000}.

\bibitem{mml}
D. F. Walls and G. J. Milburn, {\it Quantum Optics}
(Springer-Verlag, New York, 1995).

\bibitem{louisell}
W. H. Louisell {\it et al.}, \pr{124}{1646}{1961}.

\bibitem{harris}
S. E. Harris {\it et al.}, \prl{18}{732}{1967}.
\end{thebibliography}
\end{document}